\documentclass[a4paper,oneside,11pt,reqno]{amsart}
\usepackage{amssymb,bbm}
\usepackage{amsthm}
\usepackage{amsmath}
\usepackage{cite}
\usepackage{graphicx}
\usepackage[margin=2.9cm]{geometry} %Steve uses 2.7
\usepackage[foot]{amsaddr}

\usepackage{hyperref}
\hypersetup{
	pdftitle = {Algebraic units, anti-unitary symmetries, and a small catalogue of SICs}, 
	pdfauthor = {Ingemar Bengtsson}
}

\expandafter\def\expandafter\normalsize\expandafter{%
    \normalsize
    \setlength\abovedisplayskip{12pt}
    \setlength\belowdisplayskip{12pt}
    \setlength\abovedisplayshortskip{10pt}
    \setlength\belowdisplayshortskip{10pt}
}

\makeatletter
\renewcommand{\@secnumfont}{\bfseries}
\renewcommand\section{\@startsection{section}{1}%
 \z@{.7\linespacing\@plus\linespacing}{.5\linespacing}%
  {\indent\normalfont\bfseries}}
\makeatother

\begin{document}

\vspace{6cm}

\begin{center}

\

\

{\Large ALGEBRAIC UNITS, }

\

\

{\Large ANTI-UNITARY SYMMETRIES,} 

\

\

{\Large AND A SMALL CATALOGUE OF SICS}

\vspace{14mm}

{\large Ingemar Bengtsson}%$^{*}$

\vspace{14mm}

{\small {\sl  Stockholms Universitet, AlbaNova, Fysikum, \\
Stockholm, Sverige}}

\vspace{14mm}

{\bf Abstract:}

\vspace{7mm} 

\parbox{112mm}{
{\small \noindent In complex vector spaces maximal sets of equiangular lines, 
known as SICs, 
are related to real quadratic number fields in a dimension dependent way. If the 
dimension is of the form $n^2+3$ the base field has a fundamental unit of negative 
norm, and there exists a SIC with anti-unitary symmetry. We give eight examples 
of exact solutions of this kind, for which we have endeavoured to make them as 
simple as we can---as a belated reply to the referee of an earlier publication, 
who claimed that our exact solution in dimension 28 was too complicated to be fit 
to print. An interesting feature of the simplified solutions is that the components 
of the fiducial vectors largely consist of algebraic units.}} 

\end{center}

\newpage

\section{Introduction}\label{sec:intro}

\vspace{5mm}

\noindent The definition of a SIC arises naturally in quantum state 
tomography and in some corners of classical signal processing \cite{Zauner, 
Renes, FHS}. It asks for a collection of $d^2$ unit vectors forming a resolution 
of the identity in a complex vector space of dimension $d$, and such that all 
the absolute values of their mutual scalar products agree. The acronym spells 
out as ``Symmetric Informationally Complete''. All of this sounds 
simple and harmless, and at first sight it looks as if it should be easy 
to settle whether SICs exist or not. Numerical investigations 
strongly suggest that they do, in all dimensions \cite{Scott, Andrew}. 

Since 2016 it is expected that there are infinite sequences of dimensions in 
which the SICs share number theoretical properties 
\cite{AFMY}. Moreover there are ample reasons to believe that exact solutions 
for SICs would look very simple if presented in the right way. But the right 
way is expected to rely on an unwritten chapter of algebraic number theory, 
and is not yet known. This makes the SIC existence problem very interesting! 
Here we will present eight exact solutions in a sequence of dimensions of the form 
$d = n^2+3$. All of them are previously known (see Table \ref{tab:summary1}), 
but in some cases they have been available only in the form of large txt files. 
This paper is written for those who like their formulas to be confined to a 
single page. 

To see what property these dimensions share requires a look at the background. 
Zauner's conjectures \cite{Zauner}, as later refined by Appleby \cite{Marcus} 
and others, were that in every dimension $d$ SICs exist as orbits under the 
discrete Weyl--Heisenberg group, and that every such SIC is invariant under a 
symmetry of order 3. The explicit form of the symmetry is conjectured too. It 
is one of two types, $F_z$ which is realized in all dimensions, and $F_a$ 
which is realized for special SICs in dimensions of the form $d = 3\cdot (3k+1)$ 
\cite{MG2008}. Scott and Grassl  \cite{Scott, Andrew} 
conjectured that a SIC invariant under an extra anti-unitary symmetry of order 
2 exists in all dimensions of the form $d = n^2+3$, $n = 1, 2, \dots \ $. They 
also conjectured an explicit form of the symmetry. 

These conjectures have held up well as new numerical solutions have become 
available \cite{Andrew}, and do suggest that the problem has considerable depth. 
It turns out that one of our themes, that of algebraic units, is central to the 
story in more ways than one. Section \ref{sec:units} is intended to explain algebraic 
units and to sketch the ray class conjecture \cite{AFMY} for SICs. Section 
\ref{sec:simplifications} explains why SIC symmetries are especially transparent in 
certain dimensions including the ones we are concerned with. These preparations made 
we give our exact fiducial vectors in Section \ref{sec:solutions}. They are all 
located in subfields of the number field needed to construct the full SIC. However, 
writing down a vector with (say) 52 components necessarily puts a strain on typography, 
and for this reason five of the solutions are relegated to a five page 
\ref{sec:195b}ppendix. Our conclusions are in Section \ref{sec:summary}. 

Here we want to draw attention to the fact that the vector components can be 
largely (but not entirely) written in terms of algebraic units. Moreover it 
seems that these units do have a distinguished status within their respective 
unit groups. It is already known that units appear as SIC overlap phases 
\cite{AFMY}, and our results point in the same direction even though they are 
less conclusive. There are good reasons to believe that this may be quite 
important \cite{Kopp}. 

\

\section{Algebraic units}\label{sec:units}

\

\noindent Our starting point is the empirical observation \cite{AYAZ} that the 
number fields needed to write down SICs in dimension $d> 3$ are abelian extensions 
of the {\it real quadratic number field} 
${\mathbb Q}(\sqrt{D}) = {\mathbb Q}(\sqrt{D_0})$, where 

\begin{equation} D =m^2D_0 = (d+1)(d-3) \ . \label{MHG} \end{equation}

\noindent Here $m$ is an integer chosen such that $D_0$ contains no square factors. 
Recall that a quadratic number field consists of all expressions of the form 
$x + y\sqrt{D_0}$, where $x,y$ are rational numbers. Making abelian extensions 
of such number fields raises a non-trivial problem \cite{Hilbert}, but many 
insights can be obtained using eq. (\ref{MHG}) only. The first question to be 
addressed is: Given $D_0>0$, what are the possible values of $d$? 

The first step is to introduce the number $u$ through 

\begin{equation} u = \frac{d-1 + \sqrt{D}}{2} \hspace{5mm} \Leftrightarrow 
\hspace{5mm} d = 1 + u + \bar{u} \ . \label{2} \end{equation} 

\noindent The Galois conjugate $\bar{x}$ of a number $x \in {\mathbb Q}(\sqrt{D})$ 
is obtained by means of the substitution $\sqrt{D} \rightarrow - \sqrt{D}$, 
and the {\it norm} of $x$ equals $\bar{x}x$. 
Recall that a number is an {\it algebraic integer} if it is a root of a polynomial 
with integer coefficients and leading coefficient equal to $1$. An algebraic integer 
is a {\it unit} if its inverse is an algebraic integer too. It is easily checked 
that $u$ is a root of 

\begin{equation} p(t) = (t-u)(t-\bar{u}) = t^2 - (d-1)t + 1 \ . \end{equation}

\noindent Hence it is a unit of norm $\bar{u}u=+1$. A key fact about units is that 
they form a multiplicative group which is a direct product of a finite cyclic 
{\it torsion group} and a number of infinite cyclic groups. This number is the 
{\it rank} of the unit group. For real quadratic fields the rank of the unit group 
equals one, and its torsion group consists of $\pm 1$ \cite{HW}. 

It is known that the general solution of eq. (\ref{MHG}), regarded as a Diophantine 
equation for $d$ given $D_0$ with $m$ free, is the infinite sequence 

\begin{equation} d_k = 1 + u_0^k + \bar{u}_0^k \ , \end{equation}
 
\noindent where $u_0$ is known as the fundamental unit of positive norm \cite{AFMY}. 
By definition it is the smallest unit larger than 1 and having norm $+1$. 
This gives rise to a number theoretical connection between SICs in 
different dimensions. On closer investigation one finds that there are close 
geometrical ties between the SICs in such sequences, and they illuminate the 
observations about SIC symmetries that were made by Scott and Grassl \cite{Scott, 
Andrew}. In the absence of a proof of SIC existence for all $d$ the connection 
remains conjectural, but there are more than one hundred exact solutions all 
obeying this rule. 

The question now arises whether $u_0$ is a generator of the unit group. Denote the 
generator by $\eta_0$. There are two possibilities, either $u_0 = \eta_0$ or 
$u_0 = \eta_0^2$. In the latter case the generator has negative norm  
$\bar{\eta}_0\eta_0 = - 1$. A unit $\eta$ of negative norm obeys the minimal polynomial 

\begin{equation} (x-\eta)(x-\bar{\eta}) = x^2-nx - 1 = 0 \ , \end{equation}

\noindent where $n$ is some integer. Solving this equation for $\eta$ and 
squaring the result we find that 

\begin{equation} \eta = \frac{n + \sqrt{n^2+4}}{2} \hspace{5mm} 
\Rightarrow \hspace{5mm} \eta^2 = \frac{n^2+2 + n\sqrt{n^2+4}}{2} \ . \end{equation}

\noindent Comparing this with our expression for $u$ in eq. (\ref{2}) we find 
that a fundamental unit of negative norm exists if  

\begin{equation} d = n^2 + 3 \end{equation}

\noindent  for some integer $n$. This sequence in a sense `connects' several but not 
all of the sequences discussed previously. See Figure \ref{fig:sekvenser} and Table 
\ref{tab:summary1}. 

Remarkably, based on a fair amount of evidence, Scott and Grassl conjectured that in all 
these dimensions there exists a SIC with an anti-unitary symmetry of a specified form 
\cite{Scott, Andrew}. Exactly how a number theoretical property---the existence of a 
fundamental unit with negative norm---implies a geometric property---a symmetry of a 
SIC---is at the moment enigmatic.

{\tiny
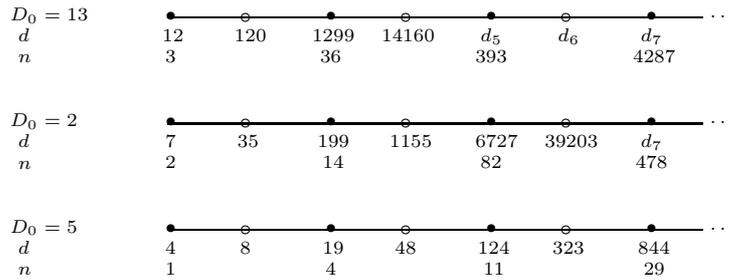
\begin{figure}[t]
\begin{picture}(350,105)
\put(32,20){$D_0=5$}
\put(35,12){$d$}
\put(35,4){$n$}
\put(91,21){\line(1,0){200}}
\put(294,21){\dots}
\put(90,20){${\bullet}$}
\put(90,12){4}
\put(90,4){1}
\put(120,21){\circle{3}}
\put(118,12){8}
\put(150,20){$\bullet$}
\put(149,12){19}
\put(150,4){4}
\put(180,21){\circle{3}}
\put(176,12){48}
\put(210,20){$\bullet$}
\put(207,12){124}
\put(209,4){11}
\put(240,21){\circle{3}}
\put(235,12){323}
\put(270,20){$\bullet$}
\put(267,12){844}
\put(269,4){29}
\put(32,60){$D_0=2$}
\put(35,52){$d$}
\put(35,44){$n$}
\put(91,61){\line(1,0){200}}
\put(294,61){\dots}
\put(90,60){$\bullet$}
\put(90,52){7}
\put(90,44){2}
\put(120,61){\circle{3}}
\put(117,52){35}
\put(150,60){$\bullet$}
\put(147,52){199}
\put(149,44){14}
\put(180,61){\circle{3}}
\put(174,52){1155}
\put(210,60){$\bullet$}
\put(206,52){6727}
\put(208,44){82}
\put(240,61){\circle{3}}
\put(232,52){39203}
\put(270,60){$\bullet$}
\put(268,52){$d_7$}
\put(266,44){478}
\put(32,100){$D_0=13$}
\put(35,92){$d$}
\put(35,84){$n$}
\put(91,101){\line(1,0){200}}
\put(294,101){\dots}
\put(90,100){$\bullet$}
\put(89,92){12}
\put(90,84){3}
\put(120,101){\circle{3}}
\put(116,92){120}
\put(150,100){$\bullet$}
\put(145,92){1299}
\put(148,84){36}
\put(180,101){\circle{3}}
\put(171,92){14160}
\put(210,100){$\bullet$}
\put(208,92){$d_5$}
\put(206,84){393}
\put(240,101){\circle{3}}
\put(237,92){$d_6$}
\put(270,100){$\bullet$}
\put(268,92){$d_7$}
\put(265,84){4287}
\end{picture}
\caption{{\small The beginnings of three infinite sequences of related 
dimensions. Dimensions with more than five digits are not written out. The 
sequence $d = n^2+3$ includes every other dimension in certain special 
sequences.}}
\label{fig:sekvenser}
\end{figure}
}

The real quadratic field is just the beginning. A much more detailed conjecture 
\cite{AFMY} concerns the full number field needed to write down SIC vectors in a 
basis preferred by the Weyl--Heisenberg group. In particular, it is believed that in 
every dimension $d$ there is a SIC that can be constructed using a {\it ray class 
field} with conductor equal to $d$ if $d$ is odd, and equal to $2d$ if $d$ is even. 
We will be concerned exclusively with such SICs here. These fields are abelian 
extensions of the real quadratic field. They have been the subject of much research 
since Hilbert formulated his 12th problem in the year 1900 \cite{Hilbert}. As a 
result these ray class fields have been completely classified, and there are 
algorithms (implemented on the computer algebra package {\sl Magma}) that will 
provide us with generators for them. However, in most cases we have to do a 
significant amount of work in order to bring the resulting set of generators 
into a useful form. 

Actually there are four relevant ray class fields, depending on whether they are 
ramified at any of the two infinite places. Whenever eq. (\ref{MHG}) relates the 
base field and the conductor the degree goes up with a factor of two for each of 
the two possible ramifications \cite{AFMY}. Number theorists refer to the largest 
ray class fields as the {\it full} ray class fields. We will simply refer to 
them as ``the ray class fields''.

In our small catalogue we have aimed to bring the ray class fields into a form that 
manifests the 
symmetry under the Galois transformation $a \leftrightarrow -a$, where $a = 
\sqrt{D_0}$. One reason for doing so is the concrete bridge between the 
SIC existence problem and the number theoretical Stark conjectures found by Kopp 
 \cite{Kopp}. This bridge rests on exactly this Galois transformation applied to 
the squared SIC overlaps, although admittedly so far only in the case that $d$ is 
a prime equal to 2 modulo 3 (as illuminated in a very recent paper \cite{Salamon}). 

We will find that algebraic units taken from the ray class fields appear in our 
exact solutions for SIC vectors. The question then arises how these units are 
placed within the unit groups of their respective fields (which will now have 
ranks considerably exceeding one). Unfortunately the available algorithms 
for calculating the generators of unit groups are very time consuming, and 
we will be able to offer only scattered comments on this question. 

{\small 
\begin{table}[h]
\caption{{\small The beginning of the sequence of dimensions considered here. The quoted 
degrees are for ray class fields ramified at both places. The Hilbert class number $h$ 
is given if it exceeds 1.}}
  \smallskip
\hskip 0.4cm
{\renewcommand{\arraystretch}{1.6}
\begin{tabular}
{|l|c|l||c|c|}\hline \hline
$d$ & $D_0$ & Degree of ray class field & Remark & Reference  \\ 
\hline
$4$ & 5 & $16=2^4$ &  & \cite{Zauner}  \\
$7$ & 2 & $24=2^3\cdot 3$ &  & \cite{Marcus} \\
$12=4\cdot 3$ & 13 & $64=2^6$ & $F_a$ & \cite{MG2008} \\
$19$ & 5 & $72=2^3\cdot 3^2$ & & \cite{Marcus} \\
$28=4\cdot 7$ & 29 & $576=2^6\cdot 3^2$ & & \cite{binom} \\
$39=3\cdot 13$ & 10 & $768=2^8\cdot 3$ & $F_a \ , \ h=2$ & \cite{ACFW}  \\
$52=4\cdot 13$ & 53 & $2304=2^8\cdot 3^2$ &  & \cite{MGun} \\
$67$ & 17 & $2904=2^3\cdot 3 \cdot 11^2$ & & \cite{MGun} \\
$84=4\cdot 3 \cdot 7$ & 85 & $4608=2^9\cdot 3^2$ & $F_a \  , \ h=2$ & \cite{MGun} \\
$103$ & 26 & $13872=2^4\cdot 3 \cdot 17^2$ & $h=2$ & \cite{MGun} \\
$124=4\cdot 31$ & 5 & $2880=2^6\cdot 3^2 \cdot 5$ & & \cite{GS} \\
$147=3\cdot 7^2$ & 37 & $4704=2^5\cdot 3 \cdot 7^2$ & $F_a$ & \cite{MGun} \\
$172=4\cdot 43$ & 173 & $28224=2^6\cdot 3^2\cdot 7^2$ & & \cite{MGun} \\
$199$ & 2 & $8712=2^3\cdot 3^2 \cdot 11^2$ & & \cite{MGun} \\
$228=4\cdot 3\cdot 19$ & 229 & $62208=2^8\cdot 3^5$ & $F_a \ , \ h=3$ & \cite{MGun} \\
$259=7 \cdot 37$ & 65 & $62208=2^8\cdot 3^5$ & $h=2$  & \cite{MGun}  \\
$292=4\cdot 73$ & 293 & $82944=2^{10}\cdot 3^4$ & & \cite{MGun} \\
$327=3\cdot 109$ & 82 & $124416=2^9\cdot 3^5$ & $F_a \ , \ h=4$ & \cite{MGun} \\
$364=4\cdot 7 \cdot 13$ & 365 & $165888=2^{11}\cdot 3^4$ & $h=2$ & \\
$403=13\cdot 31$ & 101 & $86400=2^7\cdot 3^3 \cdot 5^2$ & & \\
$444=4\cdot 3 \cdot 37$ & 445 & $331776=2^{12}\cdot 3^4$ & $F_a \ , \ h=4$ &  \\
\hline \hline
\end{tabular}
}
\label{tab:summary1}
\end{table} 
}  

\vspace{1cm}

\section{Symmetries and dimensions}\label{sec:simplifications}

\vspace{5mm}

\noindent According to Zauner's conjecture SICs are generated by a finite 
Weyl--Heisenberg group acting on some fiducial vector \cite{Zauner}. Unitary 
or anti-unitary symmetries of SIC vectors come from the Clifford group, which 
is the group of unitary and anti-unitary operators that permute the elements 
of the Weyl--Heisenberg group. It is also conjectured that every SIC 
vector has at least a symmetry of order 3, and the empirical evidence that 
this is so is by now quite overwhelming \cite{Scott, Andrew}. All the 
group theory that is relevant here is explained in a paper by Appleby 
\cite{Marcus}, except that we will also rely on the Chinese remaindering 
isomorphism that turns the Clifford group in dimension $d = n_1n_2$ into 
a direct product whenever $n_1$ and $n_2$ are relatively prime \cite{monomial}. 
From the number theoretical point of view the single most important point 
is that the Clifford group can be represented, projectively or faithfully, 
by matrices whose matrix elements belong to the {\it cyclotomic field} 
${\mathbb Q}(\tau_d)$, where 

\begin{equation} \tau_d = - e^\frac{\pi i}{d} \ . \end{equation}

\noindent The cyclotomic unit $\tau_d$ is always present in the ray class 
field containing the SIC \cite{AFMY}, but it need not be present in every fiducial 
vector. 

A point of considerable importance for us, given our aim to simplify the 
fiducial vectors as far as we can, is that whenever $d$ is a prime such that 
$d = 3$ or $d = 1$ modulo 3 the symmetry of the SIC fiducial can be implemented 
by monomial matrices. 
In particular the order 3 Zauner symmetry is then effected by a permutation 
matrix. This feature was exploited by Appleby when he gave exact solutions in 
dimensions 7 and 19, and his paper contains a full discussion of this point 
\cite{Marcus}. It was found later that if the dimension is a square then the 
entire Clifford group can be represented by monomial matrices, so that the exact 
solution in dimension 4 can be similarly simplified \cite{monomial}. We will 
find both these facts useful, because they cover all the dimensions that we 
will encounter in our sequence. 

Anti-unitary operators involve complex conjugation. When coupled with the 
preceding remarks this implies that in dimensions that are equal to a prime 
equal to 1 modulo 3, and for SICs that have an anti-unitary symmetry of order 
6, a fiducial vector can be chosen in such a way that is is real \cite{Andrew}. 
In dimension 4 special measures must be taken to achieve this \cite{monomial}. 
Here we have avoided these special measures in order to keep the action of 
the anti-unitary as transparent as possible. 

What kind of dimensions appear in our sequence? We assume that $d=n^2+3$. If $n$ 
is odd $d = 4\cdot m$ where $m$ is odd. If $n$ is even $d= m$ is odd. In both 
cases it holds either that  $m = 3\cdot k$ or that $m = k$ where $k$ is an odd 
integer equal to 1 modulo 3. Furthermore the prime decomposition of 
$k$ can contain only primes equal to 1 modulo 3. To see this, note that we can regard 
the dimension as the norm of an Eisenstein integer \cite{HW}. Recall that $\rho = 
-1/2+i\sqrt{3}/2$ is a unit in ${\mathbb Q}(\sqrt{-3})$. Thus we can write 

\begin{equation} d = n^2+3 = \left\{ \begin{array}{lll} 
(n+i\sqrt{3})\cdot (n-i\sqrt{3}) & \ & \mbox{if $n$ is even} \\ \\ 
2^2\cdot \left( \frac{n}{2} + \frac{i\sqrt{3}}{2} \right) \cdot \left( \frac{n}{2} 
- \frac{i\sqrt{3}}{2}\right) & \ & \mbox{if $n$ is odd} \ . \end{array} \right.
\end{equation}

\noindent It is known that a rational prime $p = 2$ modulo 3 remains prime considered 
as an Eisenstein integer. The class number of ${\mathbb Q}(\sqrt{-3})$ is 1. This means 
that prime factorisation is unique in this ring of integers. In particular it means that 
if an Eisenstein prime $p$ divides $d$ it must divide one of the factors in the 
decomposition we arrived at. But for a prime $p$ equal to 2 modulo 3 this is 
impossible unless $p=2$. The conclusion is that we can decompose 

\begin{equation} d = n^2+3 = 2^{a_2} \cdot 3^{a_1} \cdot p_1^{r_1}\cdot \dots \ 
\cdot p_n^{r_s} \ , \end{equation}

\noindent where $a_2\in \{ 0,2\}$, $a_1 \in \{ 0, 1\}$, and all the remaining 
primes equal 1 modulo 3.\footnote{I thank Gary McConnell for this argument.} As long 
as the multiplicities $r_i$ equal 1 
this means that the dimensions we have to deal with are of the form where we 
can use a monomial representation of the symmetry group. The ideas of ref. 
\cite{binom} may be useful in the general case, but we will not have occasion to 
use them here.

Our procedure for finding exact solutions can now be summarized as follows: We 
start with a Scott--Grassl numerical fiducial \cite{Scott} and increase the 
precision to 1200 digits. A technical point is that in the cases that the dimension 
is divisible by 3 we also make sure that the fiducial is strongly centred \cite{ACFW}. 
We then apply the Chinese remaindering isomorphism and transform to the monomial basis 
in any dimension 4 factor. In the odd dimensional factors we transform the fiducial 
so that it becomes invariant under a monomial Zauner unitary. Another technical 
point concerns SICs whose symmetry is of type $F_a$. For those we make sure that 
one of the anti-unitary symmetries is represented by pure complex conjugation. Once these 
simplifications are in place the minimal polynomials of the relative components of the 
resulting fiducial are obtained by an unsophisticated version of the method 
introduced by Appleby et al. \cite{ACFW}. The polynomials can be factored 
over a subfield of the ray class field. In fact, in the odd dimensional cases 
the cyclotomic subfield `decouples' completely from the fiducial vector, and 
enters only through the Weyl--Heisenberg group when the SIC is created from the 
fiducial. It has been noticed before that such a lowering of the degree can happen 
\cite{GS, MGun}. In the cases when the dimension is divisible by 4 we admit the cyclotomic phase 
factor $e^{i\pi/4}$ in the fiducial rather than hiding it in the monomial basis 
\cite{monomial}. The final check that we have a SIC is a straightforward 
{\sl Magma} calculation. 

The solutions, with some comments, are given on the following pages. In each case 
we obtain the $d^2$ vectors in a SIC by acting on these vectors with the 
Weyl--Heisenberg group $H(d)$ in the product form 

\begin{equation} H(p)\otimes H(3) \otimes H(4) \ , \end{equation}

\noindent where of course the three and four dimensional factors occur only 
if $d$ is divisible by 3 or 4. The representation is the standard one except 
in the four dimensional factor (if any). Thus, up to phase factors, there are 
$d^2$ group elements $X^iZ^j$. In dimension 3 the group generators $Z$ and $X$ are 
represented by the order 3 unitary matrices 

\begin{equation} Z = \left( \begin{array}{ccc} 1 & 0 & 0 \\ 
0 & \omega_3 & 0  \\ 0 & 0 & \omega_3^2 \end{array} \right) 
\hspace{12mm} X = \left( \begin{array}{rrr} 0 & 0 & 1 \\ 
1 & 0 & 0 \\ 0 & 1 & 0 \end{array} \right) \end{equation}

\noindent where $\omega_3 = e^{\frac{2\pi i}{3}}$. In the prime dimension $p$ 
we use the $p$-dimensional version of this. In dimension 4 we use the 
representation 

\begin{equation} Z = \left( \begin{array}{cccc} 0 & 0 & 1 & 0 \\ 
0 & 0 & 0 & i \\ 1 & 0 & 0 & 0 \\ 0 & i & 0 & 0 \end{array} \right) 
\hspace{9mm} X = \left( \begin{array}{rrrr} 0 & 1 & 0 & 0 \\ 
1 & 0 & 0 & 0 \\ 0 & 0 & 0 & -1 \\ 0 & 0 & 1 & 0 \end{array} \right) 
. \end{equation}

\noindent The latter is obtained from the standard representation by 
conjugation with a real unitary matrix \cite{monomial}, which is a convenient 
feature when anti-unitary transformations are of interest. Thus, although we 
do not use the standard basis in which the numerical fiducials are presented 
\cite{Scott}, in all cases the basis we do use is connected to it by means of 
a real unitary matrix whose matrix elements equal $\pm 1$ up to a harmless 
overall factor. 

\vspace{1cm}

\section{Solutions}\label{sec:solutions}

\vspace{5mm}

\noindent In this small catalogue of exact ray class SICs we arrange things so that 
the symmetry group of the fiducial is represented by monomial matrices. We use Chinese 
remaindering and the monomial basis in any dimension 4 factor. In order to make the effect 
of the Galois transformation $a \rightarrow - a$ manifest we have also tried to present 
the ray class field using generators that are either invariant or connected pairwise 
under this transformation, and such that the SIC fiducial can be expressed in terms of 
only one member of each pair. However, we failed to reach this desired form in dimension 
39. Generators that 
appear only when the field is ramified are denoted $m_1$ and $m_2$. All generators are 
real except for $m_1$. Algebraic units are denoted by the kernel letter 
$\eta$. (Please take note of this convention because we want to keep track of the units 
when they appear.) When we calculate unit groups we do so in selected subfields only. 

\

\underline{In dimension 4} the ray class field is of degree 16, and it is generated by 

\begin{equation} 
a = \sqrt{5} \hspace{9mm} r_1 = \sqrt{2} \hspace{9mm} 
m_1 = \sqrt{\frac{-a-1}{2}} \hspace{9mm} 
m_2 = \sqrt{\frac{a-1}{2}} \ . \end{equation}

\noindent We remark that $m_1m_2 = \sqrt{-1} = i$. A fiducial vector is

\begin{equation} \psi_{4a} = N\left( \begin{array}{c} 
1 \\ \sigma^7 \\ \sigma \\ x \end{array} \right) 
\hspace{10mm} 
\begin{array}{c} 
x = \frac{1}{2}(3+a)m_2 \\ \\ 
a = \sqrt{5} \hspace{10mm}
m_2 = \sqrt{\frac{a-1}{2}} \\ \\
\hspace{3mm} \sigma = \frac{1+i}{\sqrt{2}} 
= e^{\frac{i\pi}{4}} \hspace{10mm} N = \frac{1}{2}\sqrt{\frac{5-a}{5}} 
 \end{array} \label{4a}
\end{equation}

\noindent In this case the entire ray class field is used for the fiducial. 
For the Weyl--Heisenberg group we use only $\tau \in {\mathbb Q}(r_1,i)$. In 
fact $\tau = - \sigma$ in this case. The phase factor $\sigma$ will reappear 
whenever the dimension is divisible by 4. 

\

\underline{In dimension 7} the ray class field is of degree 24, and it 
is generated by

\begin{equation} \begin{array}{lll}a = \sqrt{2} & \ & \ c_1^3 - c_1^2 - 2c_1+1 = 0 
\\ \\ m_1 = \sqrt{-2a-1} & & m_2 = \sqrt{2a-1} \ . \end{array}
\end{equation} 

\noindent When a polynomial is listed as a generator we mean that one of its roots 
must be adjoined to the field. We remark that $m_1m_2 = \sqrt{-7}$ and $c_1 = 
2\cos{\frac{\pi}{7}}$. A fiducial vector is 

\begin{equation} \psi_{7b} = N\left( \begin{array}{c} 
1\\ -\eta^{-1} \\ -\eta^{-1} \\ -\eta \ \\ -\eta^{-1} \\ -\eta \ \\ -\eta \ 
\end{array} \right) 
\hspace{10mm} 
\begin{array}{c} 
\eta = \frac{1}{2}(1+a+m_2) \\ \\ 
\eta^{-1} = \frac{1}{2}(1+a-m_2) \\ \\ 
a = \sqrt{2} \hspace{12mm} m_2 = \sqrt{2a-1} \\ \\ 
N = \sqrt{\frac{3a-2}{28}} \end{array} \label{7b} 
\end{equation}

\noindent Up to a normalizing factor (that we ignore) it holds that 
$\psi_{7b} \in {\mathbb Q}(a,m_2)$ and $\tau \in {\mathbb Q}(c_1,m_1m_2)$. The 
minimal polynomial of the unit $\eta$ is 

\begin{equation} p(t) = 1 - 2 t + t^2 - 2 t^3 + t^4 \ . \end{equation}

\noindent It splits completely over ${\mathbb Q}(a,m_1,m_2)$, and defines a field whose 
unit group has rank 2. The latter is generated by $\pm 1$ together with   

\begin{equation} \eta_0 = 1 + a \ , \hspace{10mm} \eta_1= \frac{1}{2}(1-a-m_2) \ . 
\end{equation}

\noindent One of these generators is the fundamental unit in the base field. We see 
that 

\begin{equation} \eta = \eta_0\eta_1^2 \ . \end{equation}

\noindent The unit in the fiducial can be used as a generator instead of $\eta_0$. 

This solution, as well as the solution for dimension 19 below, was first 
derived in essentially this form by solving the defining equations by hand 
\cite{Marcus}. The appearance of algebraic units among the components was not 
noticed at the time. 

\

\underline{In dimension 12} the ray class field is of degree 64, and it is generated by 

\begin{equation} \begin{array}{l}a = \sqrt{13} \hspace{25mm}  r_0 = \sqrt{2} \\ \\ 
r_1 = \sqrt{\frac{5-a}{2}} \hspace{21mm} r_2 = \sqrt{\frac{5+a}{2}} \\ \\
m_1 = \sqrt{\frac{-a-1}{2}} \hspace{17mm} 
m_2 = \sqrt{\frac{a-1}{2}} \end{array} \end{equation} 

\noindent We remark that $r_1r_2 = \sqrt{3}$, $m_1m_2 = \sqrt{-3}$, and 
$\tau \in {\mathbb Q}(r_0, r_1r_2, m_1m_2)$. A fiducial 
vector is 

\begin{equation} \psi_{12b} = N\left( \begin{array}{c} 
-1 \\ \sigma^3 \\ \sigma^5 \\ r_2m_2 \\ \hline e^{i\nu_1} \\ \sigma^7e^{i\nu_1} \\ 
\sigma e^{i\nu_1} \\ e^{i\nu_2} \\ \hline e^{-i\nu_1} \\ \sigma^7e^{-i\nu_1} \\ 
\sigma e^{-i\nu_1} \\ e^{-i\nu_2}  \end{array} \right) 
\hspace{5mm} 
\begin{array}{c} 
e^{i\nu_1} = \frac{1}{8}(2m_1 + (a+1)m_2 \\ \\ 
\ \ \ + (a-1)(1-m_1m_2)) \\ \\ \\ 
e^{i\nu_2} = \frac{r_2}{24}(3(a-3)m_1 + (a-5)m_2 \\ \\ \ + (a-2)(6+2m_1m_2)) \\ \\ \\ 
\sigma = \frac{1+i}{\sqrt{2}} \hspace{10mm} N = \sqrt{\frac{13-a}{156}} \\ \end{array}
\label{12b} \end{equation}

\noindent The phase factors $e^{i\nu_1}$ and $e^{i\nu_2}$ are units. The minimal polynomial 
of $e^{i\nu_2}$ is of degree 16. It defines a field whose unit group has rank 7. Its torsion 
group is generated by $\omega = e^{2\pi i/6}$. Letting $\eta_1$ denote one out of its seven 
infinite generators we find 

\begin{equation} e^{i\nu_1} = \omega \eta_1^{-2} \hspace{8mm} e^{i\nu_2} = 
\omega^2\eta_1^3 \ . \end{equation}

\noindent Note that $e^{i\nu_1}e^{i\nu_2}$ can be used as one of the generators of this 
unit group. 

\

\underline{In dimension 19} the ray class field is of degree 72, and it is generated 
by 

\begin{equation} a = \sqrt{5} \hspace{8mm} c_1 = 2\cos{\frac{\pi}{19}} \hspace{8mm} 
m_1 = \sqrt{-2a-1} \hspace{8mm} m_2 = \sqrt{2a-1} \ . \end{equation}

\noindent We remark that $m_1m_2 = \sqrt{-19}$, and that one of the generators has 
degree nine. As seen in Figure \ref{fig:sekvenser} this dimension is the third entry 
in the sequence of dimensions starting at $d = 4$, and the ray class SIC has a 
symmetry of order 18. Still, since the fiducial vector has 19 components we feel 
forced to give it in the \ref{sec:195b}ppendix. We remark that (up to a normalizing 
factor that we can ignore) $\psi_{19e} \in 
{\mathbb Q}(a,m_2)$ and $\tau \in {\mathbb Q}(c_1,m_1m_2)$. The vector is built 
from two numbers $x_1$ and $x_2$ that are not themselves units, but the quotient $x_1/x_2$ 
is a unit. Its minimal polynomial is of degree 4, 
and the corresponding unit group has rank 2. It is in fact generated by 

\begin{equation} \eta_0 = \frac{1+a}{2} \ , \hspace{10mm} \eta_1 = 
\frac{a+5 -m_2(a+1)}{4} = \frac{x_1}{x_2} \ , \end{equation}

\noindent where $\eta_0$ is the fundamental unit of the base field and $x_1, x_2$ 
are the numbers appearing in the fiducial. 

\

\underline{In dimension 28} the ray class field is of degree $576=2^6\cdot 3^2$, 
and it is generated by 

\begin{equation} \begin{array}{lll}a = \sqrt{29} & \ & r_1 = \sqrt{2} \\ \\ 
b_1 = \sqrt{6-a} & & b_2 = \sqrt{6+a} \\ \\ 
7c_1^3 - 42 c_1 - a - 27 = 0 & &  
7c_2^3 - 42 c_2 + a - 27 = 0 \\ \\ 
m_1 = \sqrt{\frac{-a-1}{2}} & & 
m_2 = \sqrt{\frac{a-1}{2}} \end{array}
\end{equation} 

\noindent We remark that $m_1m_2 = \sqrt{-7}$. A fiducial is given in the 
\ref{sec:195b}ppendix. 
We remark that $\psi_{28c} \in {\mathbb Q}(a,b_2,c_2,m_2,\sigma)$, where the cyclotomic 
phase factor $\sigma$ enters from the dimension 4 factor only. The numbers $x_1, x_2$ 
are of degree 4, the numbers $x_3, x_4, x_5,x_6,x_7,x_8$ are of degree 24. To express 
the fiducial succinctly we have used the fact that they can be grouped into doublets 
and triplets of lower degree. The ratios of the latter six numbers are algebraic units, 
and are roots of the polynomial 

\begin{equation} p(t) = 1 - 6t - 4 t^2 + 20 t^3 + 59 t^4 - 28 t^5 + 36 t^6 - 
29 t^8 +  18 t^9 - 4 t^{11} + t^{12} \ . \end{equation}

\noindent To split this polynomial completely we need also the generators $c_1$ and $m_1$. 
The unit group of the subfield defined by it is of rank 8. Using four 
out of eight infinite generators $\eta_i$ there holds that 

\begin{eqnarray} \frac{x_3}{x_4} = \eta_1\eta_2\eta_3 \hspace{8mm} 
 \frac{x_3}{x_5} = \eta_1^2\eta_2\eta_3\eta_4^{-1} \hspace{15mm} \nonumber \\ \\ 
\frac{x_3}{x_6} = -\eta_1\eta_3 \hspace{8mm} \frac{x_3}{x_7} = 
\eta_1\eta_2\eta_3\eta_4^{-1} \hspace{8mm} 
\frac{x_3}{x_8} = -\eta_1\eta_2 \ . \nonumber \end{eqnarray}

We remark that $\tau_7 \in {\mathbb Q}(a,c_1,c_2,m_1m_2)$. The prize for adapting the 
description of the field to the fiducial vector is that the expression for the 
cyclotomic units becomes somewhat involved. 

\

\underline{In dimension 39} the ray class field is of degree $768=2^8\cdot 3$, and the 
fiducial vector in the \ref{sec:195b}ppendix belongs to a real subfield of degree 
$32 = 2^5$. There are two ray class SICs, and one can go from one to the other by 
multiplying the components with suitable units, or by a Galois transformation. It 
seems that SICs related by Galois transformations always appear when the Hilbert 
class number of the base field exceeds 1. We were not able to bring the ray class 
field into the desired form in this case (which may be due to the limitations of 
the author). Instead we resorted to using one of the relative components as one of 
the quadratic generators, which obscures the relation between the two SICs somewhat. 
The Appendix gives only one of the two fiducial vectors, and we had to resort to 
fine print to fit it into one page. Up to an overall factor, 36 of the components 
are algebraic units. 

The full ray class field can be generated by $a, r_1, b_3, b_4, m_2$ as given in 
the \ref{sec:195b}ppendix, together with 

\begin{eqnarray} r_2 = \sqrt{13} \hspace{17mm} b_1 = \sqrt{78 + 18r_2} \nonumber \\ \\ 
c^3 - 507 c + 169 = 0 \hspace{8mm} m_1 = \sqrt{-2a-1} \nonumber \end{eqnarray}

\noindent We remark that $\psi_{39i} \in {\mathbb Q}(a,r_1, b_3, b_4,m_2)$, 
$\tau_3\in {\mathbb Q}(r_2m_1m_2)$, and $\tau_{13} \in 
{\mathbb Q}(r_2,b_1,c,m_1m_2)$. Note that ramification at one infinite place is needed 
to define the real degree four generator $b_4$. The fiducial is built from fourteen 
numbers $x_1 , \dots, x_{14}$, and one finds that the twelve ratios $x_3/x_1, \dots , 
x_{12}/x_1$ are units. 

\

\underline{In dimension 52} the ray class field is of degree $2304 = 2^8\cdot 3^2$, 
and it is generated by 

\begin{equation} \begin{array}{l}a = \sqrt{53} \hspace{29mm}  r_1 = \sqrt{2} \\ \\ 
b_1 = \sqrt{15+2a} \hspace{18mm} b_3 = \sqrt{\frac{13+(a-8)b_1}{2}} \\ \\ 
b_2 = \sqrt{15-2a} \hspace{18mm} b_4 = \sqrt{\frac{13-(a+8)b_2}{2}} \\ \\ 
2c_1^3 + (3-a)c_1^2 - (7+a)c_1 + 4a - 6 = 0 \\ \\ 
2c_2^3 + (3+a)c_2^2 - (7-a)c_2 - 4a - 6 = 0 \\ \\ 
m_1 = \frac{1}{13}\sqrt{34-5a} \hspace{13mm} 
m_2 = \frac{1}{13}\sqrt{34+5a} \end{array} \end{equation} 

\noindent We remark that $m_1m_2 = i$. A fiducial vector is given in the 
\ref{sec:195b}ppendix, and we have to admit to the use of some fine print there. We remark that 
$\psi_{52d} \in {\mathbb Q}(a,b_2,b_4,c_2,m_2, \sigma)$. Except for $m_2, x_1, x_2, x_3, x_4$ 
all the numbers appearing in the fiducial are units. We did not investigate the unit 
group. Finally $\tau_{13} \in {\mathbb Q}(a,c_1,c_2,, b_1b_2, b_3b_4, m_1m_2)$. So again 
our choices have complicated the description of the Weyl--Heisenberg group. 

\

\underline{Dimension 67} exhibits the curious feature that there are two inequivalent 
orbits (67a and 67b in Scott's notation \cite{Andrew}) that have anti-unitary symmetry, 
and yet the class number of the base field equals 1 so only one ray class SIC is 
expected. However, the degree of the ray class field field includes a factor 
of $11^2$, and considerable skill is needed to analyze this dimension \cite{MGun}. 

\

We jump to an irresistible example in a higher dimension. 
\underline{In dimension 124} we encounter the fifth entry in the dimension 
sequence that starts at $d = 4$, and the ray class SIC has a symmetry of order 
30. Because 4 divides 124 the ray class field contains the $d = 4$ ray class field as a 
subfield \cite{AFMY}. We give a fiducial in the \ref{sec:195b}ppendix, without using 
fine print. Except for the number 2, and the normalizing factor, all the numbers that 
enter are units. The full ray class field can be generated by $a, b_2, c_2, m_2$ 
as given in the \ref{sec:195b}ppendix, together with 

\begin{eqnarray} r_1 = \sqrt{2} \hspace{10mm} b_1 = \sqrt{6-a} \hspace{10mm} 
m_1 = \sqrt{\frac{-a-1}{2}} \hspace{7mm} \nonumber \\ \nonumber \\ 
2c_1^3 - (13 +7a)c_1^2 + (51+23a)c_1 - 18 - 8a = 0 \hspace{7mm} \\ \nonumber \\ 
y^5 - 4805y^3 + 4617605y + \frac{1}{2}(3723875a - 21419729) = 0 
\nonumber \end{eqnarray}

\noindent To build the components of $\psi_{124a}$ we need a subfield of the 
modest degree 24, supplemented by the eigth root of unity $\sigma$ coming from 
the way we handle the dimension four factor. The 
quintic polynomial can be brought into a form more in conformity with 
our wishes at the expense of some further work. It must be admitted that the 
cyclotomic phase factor $\tau_{31} \in {\mathbb Q}(a,b_1b_2m_1m_2, c_1c_2)$ now 
looks quite atrocious, and the final check of the SIC condition is time consuming.

\vspace{1cm}

\section{Conclusions}\label{sec:summary}

\vspace{5mm}

\begin{figure}[h]
\begin{center}
\includegraphics[width=.3\textwidth]{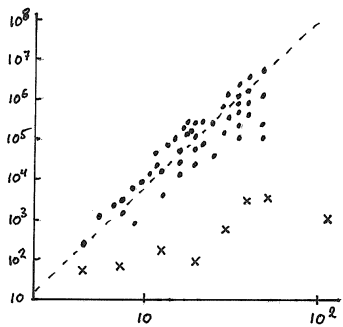}
\end{center}
\caption{\small The number of characters needed to describe the fiducial vectors given in refs. 
\cite{Scott} and \cite{ACFW} (dots), and the solutions given here (crosses). The dimension 
is given on the abscissa and the number of characters on the ordinate, in logarithmic 
scales. The dashed line is a `prediction' by Magsino and Mixon \cite{MMixon}, who plotted the dots.}
\label{fig:Mixon}
\end{figure}

%Antalet tecken som ger rättningsvektorn i txt-filerna:
% 4: 51, 7: 62, 12: 184, 19: 105, 28: 611, 39 3160, 52: 3370, 124:1270

\noindent The apparent complexity of the exact solutions for SICs has been noticeable 
from the start. Magsino and Mixon had the amusing idea to illustrate this by plotting 
the number of characters used to describe the coordinates of the fiducial vectors 
against the dimension \cite{MMixon}. By adding our solutions to their plot we see 
clearly that the description used here is very economical at least in this naive sense, 
see Figure \ref{fig:Mixon}. This is achieved by locating the fiducial vector in a subfield 
that is `complementary' to the cyclotomic subfield of the ray class field. While our simplified 
solutions do not have the elegance achieved for the dimension ladder $d = 5, 15, 195$ 
\cite{SiSi}, the fact that we were able to manifest the behaviour under the Galois 
transformation $a \rightarrow - a$ (in all but one case) is some compensation.

The most striking feature seems to be the appearance of algebraic units among the 
components of the vectors (in all cases). Moreover our admittedly superficial analysis 
suggests that they are quite distinguished units. This observation may serve as a 
point of departure for further studies. 

\vspace{1cm}

\noindent \underline{Acknowledgements}: I thank Gary McConnell for help with (and 
patient explanations of) the number theory, Marcus Appleby for (indispensable) 
help with the calculations, and Markus Grassl for comments on the manuscript and 
for insights into dimension 67. I also thank the Mathematics Department at Stockholm 
University for allowing me to use their computer. 

\vspace{1cm}

\appendix

\section{Exact solutions}\label{sec:195b}

\

\noindent Here we give exact fiducial vectors for the SIC orbits 
19e, 28c, 39i, 52d, and 124a. Exact fiducials for 4a, 7b, and 
12b are given in the main text, see eqs. (\ref{4a}), (\ref{7b}), and (\ref{12b}). 
All our examples are available in exact form elsewhere, see Table \ref{tab:summary1} 
for references. Numbers denoted with the kernel letter $\eta$ are algebraic units. 
There are detailed comments on the solutions in the main text. 

\vspace{1cm} 

\begin{equation} \psi_{19e} = N\left( \begin{array}{r} 
1 \ \\ -x_1 \\ x_2 \\ x_2 \\ -x_1 \\ -x_1 \\ -x_1 \\ -x_1 \\ x_2 \\ -x_1 \\ x_2 
\\ -x_1 \\ x_2 \\ x_2 \\ x_2 \\ x_2 \\ -x_1 \\ -x_1 \\ x_2 \end{array} \right) 
\hspace{10mm} 
\begin{array}{c} 
x_1 = \frac{1}{10}a(m_2-1) \\ \\ 
x_2 = \frac{1}{10}a(m_2+1) \\ \\ \\ 
\eta_1 = \frac{x_1}{x_2} = \frac{1}{4}(a+5 - m_2(a+1)) \\ \\ \\ 
a = \sqrt{5} \\ \\ 
m_2 = \sqrt{2a-1} \\ \\ \\ 
N = \sqrt{\frac{9a-5}{76}} \end{array}
\end{equation}

\newpage

\begin{eqnarray} \psi_{28c} = N\left( \begin{array}{c} 
 0 \\ 0 \\ 0 \\ 2 \\\hline x_3\sigma^6 \\ x_4\sigma \\ x_5\sigma^7 \\ x_1\sigma^4 \\ 
 \hline  x_5\sigma^6 \\ x_3\sigma^5 \\ x_4\sigma^3 \\ x_1\sigma^4 \\ 
 \hline x_6\sigma^2 \\ x_7\sigma^5 \\ x_{8}\sigma^3 \\ x_2\sigma^4 \\ \hline
 x_4\sigma^2 \\ x_5\sigma^5 \\ x_3\sigma^7 \\ x_1\sigma^4 \\ \hline 
 x_7\sigma^6 \\ x_{8}\sigma \\ x_6\sigma^3 \\ x_2\sigma^4 \\ \hline 
x_8\sigma^2 \\ x_6\sigma \\ x_7\sigma^7 \\ x_2\sigma^4 \end{array} \right) 
\hspace{10mm} 
\begin{array}{c} 
x_1 = \frac{1}{4}(a+5 - 2m_2) \\ \\ 
x_2 = \frac{1}{4}(a+5+2m_2) \\ \\ 
x_3 = \frac{b_2}{36}((-4c_2^2 - (a - 9)c_2 + 10b_2)m_2 \\ \hspace{3mm} + 6c_2 + 
3a - 9) \\ \\ 
x_3x_4 = \frac{1}{36}(((a + 13)c_2^2 + (a - 1)c_2 - a - 37)m_2 \\ 
+ (a - 1)c_2^2 - 2(2a + 5)c_2 - a + 1) \\ \\ 
x_3x_8 = \frac{1}{18}(-(a - 1)c_2^2 + 2(2a + 5)c_2 + a - 1) \\ \\ 
x_4x_6 = \frac{1}{18}((a + 13)c_2^2 + (a - 1)c_2 - a - 55) \\ \\ 
x_3x_4x_5 = x_6x_7x_8 = b_2 \\ \\ \\ 
\eta_1 = \frac{x_7}{x_5} \hspace{6mm} \eta_2 = -\frac{x_6}{x_4} \hspace{6mm} 
\eta_3 = -\frac{x_8}{x_4} \hspace{6mm} \eta_4 = \frac{x_7}{x_4} \\ \\ \\ 
\sigma = e^\frac{i\pi}{4} \hspace{8mm} N = \sqrt{\frac{12a-58}{812}} \\ \\ \\ 
a = \sqrt{29} \hspace{10mm} b_2 = \sqrt{a+6} \\ \\ 7c_2^3 - 42 c_2 + a - 27 = 0 \\ \\ 
m_2 = \sqrt{\frac{a-1}{2}} \end{array} 
\nonumber \end{eqnarray}

\newpage

{\tiny \begin{eqnarray} 
%{\tiny 
\begin{array}{l} 
\psi_{39i} = N  \left( \begin{array}{r} 
 1 \ \\ -x_1 \\ x_2 \\ \hline 
- x_3 \\ x_7 \\ - x_8 \\ \hline 
x_5 \\ - x_9 \\ x_{10} \\ \hline - x_3 \\ x_7 \\ -x_{8} \\ \hline 
x_4 \\ x_{11} \\ x_{12} \\ \hline x_5 \\ -x_9 \\ x_{10} \\ \hline 
x_5 \\ -x_9 \\ x_{10} \\ \hline x_6 \\ x_{13} \\ x_{14} \\ \hline 
x_6 \\ x_{13} \\ x_{14} \\ \hline -x_3 \\ x_7 \\ -x_{8} \\ \hline 
x_4 \\ x_{11} \\ x_{12} \\ \hline x_6 \\ x_{13} \\ x_{14} \\ \hline 
x_4 \\ x_{11} \\ x_{12} \end{array} \right)  \\ \\ 
\\ \\ \\ 
N = \sqrt{\frac{6(51a-150)-(23a-86)r_1}{780}} \\ \\ \\ \\ \\ 
\eta_1 = \frac{x_3}{x_1} \\ 
\ \ \ \ \ \vdots \\ \\ 
\eta_{12} = \frac{x_{14}}{x_1} 
\end{array} %} 
\hspace{1mm} 
%{\tiny 
\begin{array}{c} 
x_1 = -\frac{1}{520}(r1 - 10)b_3m_2 - \frac{1}{120}(7a + 25)r_1 + \frac{1}{4}(a + 3) \\ \\ 
x_1x_2 = \frac{1}{60}(13a + 40)r_1 - \frac{1}{18}(11a + 38) \\ \\ 
x_3 = b_4 \\ \\ 
x_5 = \frac{1}{1520}(1430 + 650 a + (-195 - 117 a) r_1 + b_3 (-250 - 46 a + (53 + 11 a) r_1) \\ 
+ b_4 (390 - 390 a + (195 + 117 a) r_1 + b_3 (210 - 30 a + (-63 + 9 a) r_1))) \\ \\ 
x_3x_4 = \frac{1}{2340}(((-22a - 67)r_1 + 6(17a + 50))b_3 + 130(a + 1)) \\ \\ 
x_5x_6 = \frac{1}{2340}(((-22a - 67)r_1 + 6(17a + 50))b_3 - 130(a + 1)) \\ \\

x_7 = \frac{1}{3120}(-2340 - 1170 a + (650 + 299 a) r_1 + b_3 (400 + 58 a - (92 + 11 a) r_1) \\ 
+ b_4 (-390 + 390 a + (195 - 273 a) r_1 + b_3 (-390 + 78 a + (63 - 9 a) r_1)) \\ 
+  m_2 (1560 + 390 a + (-260 - 65 a) r_1 + b_3 (-100 - 34 a + (22 + 7 a) r_1) \\ 
+ b_4 (-390 + (195 + 78 a) r_1 + b_3 (-30 + 24 a + (3 - 6 a) r_1)))) \\ \\

x_9= \frac{1}{3120}(-650 - 260 a + b_3 (150 + 12 a - 39 r_1) - 195 r_1 \\ 
+ b_4 (-1170 + 390 a - (195 + 39 a) r_1 + b_3 (-330 + 114 a + (69 - 21 a) r_1)) \\ 
+ m_2 (910 + 130 a - (195 + 39 a) r_1 + b_3 (-50 - 38 a + (9 + 9 a) r_1) \\ 
+ b_4 (-390 + 195 r_1 + b_3 (-30 + 36 a + (3 - 6 a) r_1)))) \\ \\

x_{11} = \frac{1}{3120}(2080 + 910 a + (-260 - 143 a) r_1 + b_3 (-280 + 14 a + (60 - 3 a) r_1) \\ 
+  b_4 (390 - 390 a + (-195 + 273 a) r_1 + b_3 (390 - 78 a + (-63 + 9 a) r_1)) \\ 
+ m_2 (-1820 - 650 a + (390 + 117 a) r_1 + b_3 (60 + 30 a - (10 + 7 a) r_1) \\ 
+ b_4 (390 - (195 + 78 a) r_1 + b_3 (30 - 24 a + (-3 + 6 a) r_1)))) \\ \\ 

x_{13} =\frac{1}{3120}(-910 - 520 a + (195 + 156 a) r_1 + b_3 (30 - 60 a + (-7 + 14 a) r_1) \\ 
+ b_4 (-1170 + 390 a + (-195 - 39 a) r_1 + b_3 (-330 + 114 a + (69 - 21 a) r_1)) \\ 
+ m_2 (1170 + 390 a + (-325 - 91 a) r_1 + b_3 (-90 - 42 a + (21 + 9 a) r_1) \\ 
+ b_4 (-390 + 195 r_1 + b_3 (-30 + 36 a + (3 - 6 a) r_1)))) \\ \\ 

x_7x_8 = \frac{1}{4680}(-23790 - 7410 a + (5395 + 1651 a) r_1 \\ 
+ b_3 (1850 + 590 a - (421 + 133 a) r_1) \\ 
+ b_4 (10140 + 3120 a + (-1950 - 546 a) r_1 \\ 
+ b_3 (-360 - 216 a + (96 + 42 a) r_1)))\\ \\ 

x_9x_{10}  = \frac{1}{4680}(-10010 - 2990 a + (2015 + 611 a) r_1 \\ 
+ b_3 (690 + 258 a - (147 + 57 a) r_1) \\ 
+ b_4 (1950 + 780 a + (-975 - 156 a) r_1 \\ 
+ b_3 (-390 - 156 a + (99 + 36 a) r_1))) \\ \\ 

x_7x_8x_{11}x_{12} = \frac{1}{21060}(226200 + 71760 a - (50635 + 16003 a) r_1 \\ 
+ b_3 (-13090 - 4162 a + (2928 + 930 a) r_1)) \\ \\ 

x_9x_{10}x_{13}x_{14} = \frac{1}{1/21060}(226200 + 71760 a - (50635 + 16003 a) r_1 \\ 
+ b_3 (13090 + 4162 a + (-2928 - 930 a) r_1))\\ \\ \\ 

a = \sqrt{10} \hspace{5mm} r_1 = \sqrt{20} \hspace{5mm} b_3 = \sqrt{13(4a+15)(r_1+5)} \hspace{5mm} m_2 = \sqrt{2a-1} \\ \\ 
b_4^2 - \frac{1}{1560}(((6a - 3)r_1 - 20a + 10)b_3 - 13((8a + 35)r_1 - 40a - 130))b_4 \\ 
+ \frac{1}{2340}((22a + 67)r_1 - 6(17a + 50))b_3 - \frac{1}{18}(a + 1) = 0 
  \end{array} %}
\nonumber \end{eqnarray} }

\newpage

{\tiny \begin{eqnarray} %{\tiny 
\begin{array}{c} \psi_{52d} = %{\tiny 
N\left( \begin{array}{c} 
\sigma^2 \\ \sigma \\ \sigma^3 \\ m_2^{-1} \\ 
\hline \eta_1 \sigma^2 \\ \eta_2\sigma \\ \eta_3\sigma^7 \\ x_1\sigma^4 \\ 
\hline \eta_4\sigma^2 \\ \eta_5\sigma^5 \\ \eta_6\sigma^3 \\ x_2 \\ 
\hline \eta_2\sigma^2 \\ \eta_3\sigma^5 \\ \eta_1\sigma^3 \\ x_1\sigma^4 \\ \hline
\eta_7\sigma^6 \\ \eta_8\sigma \\ \eta_9\sigma^7 \\ x_3 \\ \hline 
\eta_6\sigma^2 \\ \eta_4\sigma^1 \\ \eta_5\sigma^7 \\ x_2 \\ \hline 
\eta_5\sigma^6 \\ \eta_6\sigma \\ \eta_4\sigma^3 \\ x_2 \\ \hline 
\eta_{10}\sigma^2 \\ \eta_{11}\sigma\\ \eta_{12}\sigma^3 \\ x_4 \\ \hline 
\eta_{11}\sigma^2 \\ \eta_{12}\sigma \\ \eta_{10}\sigma^3 \\ x_4 \\ \hline 
\eta_3\sigma^6 \\ \eta_1\sigma \\ \eta_2\sigma^3 \\ x_1\sigma^4 \\ \hline 
\eta_9\sigma^6 \\ \eta_7\sigma^5 \\ \eta_8\sigma^3 \\ x_3 \\ \hline 
\eta_{12}\sigma^2 \\ \eta_{10}\sigma \\ \eta_{11}\sigma^3 \\ x_4 \\ \hline 
\eta_8\sigma^2 \\ \eta_9\sigma^5 \\ \eta_7\sigma^7 \\ x_3 \end{array} \right) %} 
\end{array} %}
\hspace{2mm} 
%{\tiny 
\begin{array}{c} 
x_1 = \frac{m_2}{8}((a + 1)b_2 - a - 1) + \frac{1}{28}((a + 2)b_2 + 2a - 3)b_4 \\ \\ 
x_2 = \frac{m_2}{8}((a + 1)b_2 + a + 1) - \frac{1}{28}((a + 9)b_2 + 2a - 10)b_4 \\ \\ 
x_1x_3 = \frac{1}{2}(7-a) + b_2 \hspace{12mm} x_2x_4 = \frac{1}{2}(a - 7) + b_2 
\\ \\ \\ 

\eta_1 = \frac{m_2b_4}{1820}(((22a - 82)b_2 - 26(a - 12))c_2^2 - 
((46a - 279)b_2 + 13(20a - 149))c_2 \\ + 
\frac{91}{2}((a + 1)b_2 + 5a - 47)) - 
 \frac{1}{520}(((6a + 32)b_2 - 26a + 104)c_2^2 \\ + ((8a + 34)b_2 - 26a)c_2 
-(19a + 97)b_2 + 13(11a - 55)) \\ \\ 
\eta_2 = \frac{m_2b_4}{3640}((-(41a - 583)b_2 - 13(23a - 185))c_2^2 + 
((87a + 321)b_2 + 13(43a - 243))c_2 \\ + 2(327a - 1649)b_2 + 26(53a - 363)) - 
\frac{1}{520}(((2 a + 2)b_2 + 52)c_2^2 \\ 
+ ((7a + 85)b_2 + 13(3a - 19))c_2 + 4(6a + 45)b_2 - 52 a - 104) \\ \\ 
\eta_1\eta_2\eta_3 = \frac{m_2b_4}{28}((19a - 137)b_2 + 2(19a - 137)) + 
 \frac{1}{8}((a - 7)b_2 - 3a + 21) 
\\ \\ \\
\eta_4 = \frac{m_2b_4}{3640} ((5(11a - 41)b_2 - 13(19a - 137))c_2^2 
- (10(2a - 115)b_2 + 26(a - 12))c_2 \\ 
- 10(11a - 41)b_2 + 26(103a - 781)) + 
 \frac{1}{520}(((2a + 2)b_2 - 52)c_2^2 \\ 
+ ((7a + 85)b_2 - 13(3a - 19))c_2 + 
4(6a + 45)b_2 + 52a + 104) \\ \\ 
\eta_5 = \frac{m_2b_4}{3640}(((2(53a - 90)b_2 - 26(a - 12))c_2^2 + 
((79a + 781)b_2 - 13(25a - 209))c_2 \\ 
- 2(341a - 1271)b_2 - 26(25a - 209)) + \frac{1}{520} (((8a + 34)b_2 + 26(a - 6))c_2^2 \\ 
+ ((15a + 119)b_2 - 13(a - 19))c_2 - (60a + 242)b_2 - 26(6a - 39)) \\ \\ 
\eta_4\eta_5\eta_6 = \frac{m_2b_4}{28}((-a + 12)b_2 + 96a - 697) - 
\frac{1}{8}((a - 7)b_2 + 3a - 21)
\\ \\ \\
\eta_8\eta_1 = -\frac{1}{260}(((a + 79)b_2 + 13(3a - 23))c_2^2 \\ + 
((-3a + 127)b_2 + 13(a - 9))c_2 - (a + 391)b_2 - 13(3a - 27)) \\ \\ 
\eta_9\eta_2 = \frac{1}{260}((-4 a + 48)b_2c_2^2 \\ 
+ ((25a - 27)b_2 + 65(-a + 7))c_2 + (30a + 56)b_2 + 130(2a - 15)) \\ \\ 
\eta_1\eta_2\eta_3\eta_7\eta_8\eta_9 = 25a-182 
\\ \\ \\
\eta_{1}\eta_8\eta_6\eta_{11} = 
\frac{1}{10}((-17a + 121)c_2^2 - (11a - 77)c_2 + 43a - 301) \\ \\ 
\eta_{3}\eta_7\eta_5\eta_{10} = 
\frac{1}{10}((-20a + 142)c_2^2 + (19a - 147)c_2 + 183a - 1309) \\ \\ 
\eta_4\eta_5\eta_6\eta_{10}\eta_{11}\eta_{12} = 25a-182 
\\ \\ \\
\sigma = e^\frac{i\pi}{4} \hspace{8mm} 
N = \frac{1}{2}\sqrt{\frac{53+a}{689}} \\ \\ \\ 
a = \sqrt{53} \hspace{8mm} b_2 =\sqrt{15-2a} \hspace{8mm} 
b_4 = \sqrt{\frac{13-(a+8)b_2}{2}} \\ \\ 
2c_2^3 + (3+a)c_2^2 - (7-a)c_2 - 4a - 6 = 0 \\ \\  
m_2 = \frac{1}{13}\sqrt{34+5a}  
 \end{array}
%}
\nonumber \end{eqnarray} }

\newpage

\begin{eqnarray}
\psi_{124a} =N\left( \begin{array}{c} 
 {\bf w}_0 \\ {\bf u}_1 \\ {\bf u}_1 \\ {\bf v}_1 \\ {\bf u}_1 \\ 
{\bf u}_2 \\ {\bf v}_1 \\ {\bf u}_3 \\ {\bf u}_1 \\ 
{\bf u}_2 \\ {\bf u}_{2} \\ {\bf v}_{2} \\ {\bf v}_{1} \\ 
{\bf v}_{2} \\ {\bf u}_{3} \\ {\bf v}_{3} \\ {\bf u}_{1} \\ 
{\bf v}_{1} \\ {\bf u}_{2} \\ {\bf u}_{3} \\ {\bf u}_{2} \\ 
{\bf v}_{2} \\ {\bf v}_{2} \\ {\bf v}_{3} \\ {\bf v}_{1} \\ 
{\bf u}_{3} \\ {\bf v}_{2} \\ {\bf v}_{3} \\ {\bf u}_{3} \\ 
{\bf v}_{3} \\ {\bf v}_{3} \\ \end{array} \right) 
\hspace{1mm} 
\begin{array}{c} 
\eta_1 = \frac{1}{2}(3-a)m_2 \\ \\ 
\eta_2  = \frac{1}{4}((a-1)b_2-(a+1)m_2) \hspace{8mm} 2\eta_2\eta_3 = 3 - a \\ \\ 
\eta_4 = \frac{1}{124}((-(11a + 27)c_2^2 - (5a + 1)c_2 + 10a + 2)b_2m_2 \\ 
+ 31(a + 3)c_2^2 + 31(3a + 1)c_2) \\ \\ 
\eta_5 = \frac{1}{124}(((38a + 82)c_2^2 + (37a + 119)c_2 + 19a - 83)b_2m_2 \\ - 62c_2 - 62(a - 2)) \\ \\ 
\eta_7 = \frac{1}{124}(((38a + 82)c_2^2 + (37a + 119)c_2 + 19a - 83)b_2m_2 
\\ + 62c_2 + 62(a - 2)) \\ \\ 
\eta_9 = \frac{1}{124}(((11a + 27)c_2^2 + (5a + 1)c_2 - 10a - 2)b_2m_2 \\ 
+ 31(a + 3)c_2^2 + 31(3a + 1)c_2) \\ \\ 
\eta_4\eta_5\eta_6 = \eta_7\eta_8\eta_9 = a - 2 \\ \\ \\ 
\sigma = e^\frac{i\pi}{4} \hspace{10mm} N = \frac{1}{10}\sqrt{\frac{25+a}{31}} \\ \\ \\ 
a = \sqrt{5} \hspace{8mm} b_2 = \sqrt{a+6} \hspace{8mm} m_2 = \sqrt{\frac{a-1}{2}} 
\\ \\ 
2c_2^3 -(13-7a)c_2^2 + (51-23a)c_2 - 18 +8a = 0  
\end{array} \nonumber \end{eqnarray}

\begin{eqnarray} 
{\bf w}_0 = 
\left[ \begin{array}{l} 2 \\ 2\sigma^7 \\ 2\sigma \\ 2\eta_1 \end{array}  \right] 
\hspace{4mm} {\bf u}_{1} = 
\left[ \begin{array}{l} \eta_4 \\ \eta_5\sigma^7 \\ \eta_6\sigma^5 \\ 
\eta_2 \end{array}  \right] 
\hspace{4mm} {\bf u}_{2} = 
\left[ \begin{array}{l} \eta_6\sigma^4 \\ \eta_4\sigma^7 \\ \eta_5\sigma \\ 
\eta_2 \end{array} \right] 
\hspace{4mm} {\bf u}_{3} = 
\left[ \begin{array}{l} \eta_5 \\ \eta_6\sigma^3 \\ \eta_4\sigma \\ 
\eta_2 \end{array} \right] 
\nonumber \\ \nonumber \\ \nonumber \\ 
{\bf v}_{1} = 
\left[ \begin{array}{l} \eta_7\sigma^4 \\ \eta_8\sigma^7 \\ \eta_9\sigma \\ 
\eta_3\sigma^4 \end{array} \right] 
\hspace{9mm} {\bf v}_{2} = 
\left[ \begin{array}{l} \eta_8 \\ \eta_9\sigma^7 \\ \eta_7\sigma^5 \\ 
\eta_3\sigma^4 \end{array} \right] 
\hspace{9mm} {\bf v}_{3} = 
\left[ \begin{array}{l} \eta_9 \\ \eta_7\sigma^3 \\ \eta_8\sigma \\ 
\eta_3\sigma^4 \end{array} \right] \hspace{8mm} 
\nonumber \end{eqnarray}

\end{document}